\documentclass{emulateapj}

\shorttitle{Radial Alignment of Cluster Galaxies}
\shortauthors{Pereira and Kuhn}

\begin{document}

\title{Radial Alignment of Cluster Galaxies}

\author{M. J. Pereira and J. R. Kuhn}

\affil{Institute for Astronomy, University of Hawaii, Honolulu-HI-96822}

\begin{abstract}

We report the discovery of a statistically significant radial alignment of cluster galaxies in a sample of 85 X-ray selected clusters observed in the Sloan Digital Sky Survey. The tendency for alignment is a robust feature of our sample and does not vary significantly with individual cluster  or galaxy properties. We use dynamical arguments to show that a significant fraction of cluster galaxies should be undergoing a parametric tidal resonance that can cause their long axes to orient themselves towards the center of the cluster potential, and therefore tentatively ascribe the observed radial alignment to this dynamical effect. 




\end{abstract}


\section{Introduction}

The search for large-scale systematic galaxy alignments and their physical causes goes back many years. Even before sensitive large-scale photographic surveys existed, Brown
(1938) suggested that galaxy orientations are not isotropic. Hawley and
Peebles (1975) were probably the first to analyze a large modern galaxy
database, finding statistically weak evidence that Coma cluster
galaxies are preferentially oriented toward the cluster center. To our knowledge, Coma is the only system where any evidence for radial alignment of cluster galaxies has been found (Thompson 1976; Djorgovski 1983). On larger scales, galaxy 
alignments have been seen with respect to the cluster major-axis and with respect to neighboring clusters
(e.g.\ Bingelli 1982;  Plionis et al. 2003). Galaxy cluster data generally
support the view that clusters and their brightest galaxies do tend to
align (c.f.\ Plionis et al. 2003, Struble 1990) although alignment of a
significant population of cluster galaxies with the cluster axis or
toward the cluster center has not yet been established (i.e. 
Fong, Stevenson, Shanks 1990; Trevese,
Cirimele, and Flin 1992).


Much work has been done towards formulating plausible physical mechanisms that could cause alignments in cluster galaxies and between clusters. Hawley and Peebles (1975) reviewed some of the early speculation about how fossil eddy turbulence or primeval magnetic fields might be a cause for such anisotropy. Binney and Silk (1979) suggested that tidal interactions between protostructures could cause anisotropies in galaxy orientations, and West (1994) suggests that anisotropic mergers along large-scale filaments could induce an alignment of the BCGs with the cluster axis and neighbouring large-scale structure. Catelan \& Porciani (2001) describe how tidal interactions can generate a correlation between the direction of the angular-momentum vector of a subhalo and the surrounding tidal shear field. It is important to note that all these mechanisms are primordial, and one would expect the alignment they induce at formation to decay with galaxy-cluster and galaxy-galaxy interactions. A particularly interesting question then emerges: is galaxy alignment a feature of the initial conditions of cluster formation, as assumed in virtually all of the existing literature,  or is it the result of a dynamical effect that grows with galaxy and cluster evolution? These different scenarios lead to distinct alignment tendencies (i.e.\ radial vs cluster-axis vs cluster-cluster alignments) and redshift evolution. An extensive survey of the strength of galaxy alignments in clusters and its correlation with individual properties of the parent systems such as morphology and redshift should help to determine which (if any) is the dominant mechanism.

This letter reports the first detection of a statistically significant {\it radial} alignment of cluster galaxies for a homogeneously selected sample of clusters. In Section 2 we define our sample and describe the data extraction and membership selection procedures. Section 3 presents our findings for the resulting samples of galaxies and addresses possible systematic errors in our measurement. In Section 4 we describe the dynamical tidal interaction between a galaxy and the gravitational potential of its host cluster and how it can lead to a statistical tendency for radial galaxy alignment. The final section comprises a brief discussion of these results  and a look into the future of galaxy alignment studies. 

\section{A Large Sample of Cluster Galaxies}

\subsection{Sample Definition}

Our data are selected using the largest X-ray and optical cluster
samples currently available in the northern hemisphere. We obtain our cluster
targets from the Extended Brightest Cluster Sample
(eBCS, Ebeling et al., 1998, 2000), a $90\%$ complete, X-ray flux-limited sample of
the brightest clusters in the northern hemisphere, compiled from the
ROSAT All-Sky Survey (RASS) data.  By limiting our sample to known
X-ray luminous clusters we avoid spurious cluster identifications from
projection effects and ensure that we are looking at fairly massive
systems. We cross-correlate the eBCS list with the Sloan Digital Sky Survey (SDSS) "footprint'' from
data release 3 (Abazajian et al. 2005) and find that, out of 301 eBCS clusters, 108 have been observed by the SDSS so far.


\subsection{Data Extraction}

For the data extraction, we define a preliminary circular cluster region, 2 Mpc in radius, centered on the
X-ray position, and a background annulus, 1 Mpc wide, 4-5 Mpc away from the same point. We assume that this is far
enough for the background region to be unaffected by the presence of
the cluster but, for most clusters, close enough for it to suffer the same
sort of systematic uncertainties intrinsic to the survey,
e.g. Galactic extinction. We assume throughout a flat,
$\Lambda$-dominated universe, with $\Omega_{0}=0.3$, $\Omega_{\Lambda}
= 0.7$ and $h = 0.75$. Photometric and, where available, spectroscopic
data are extracted for all galaxies that fall within these regions and for
which position angles have been determined in the SDSS.

\subsection{Cluster Membership}
	
The most practical and accurate way to determine which galaxies are
gravitationally bound to each cluster is to get 3D information directly
from spectroscopically determined redshifts.  Any object with a
velocity within $3 \sigma$ (where $\sigma$ is the cluster velocity dispersion) of the systemic velocity of the cluster that is also within 2 Mpc projected distance
of its center is selected as a cluster member (sample A). We use a robust
statistical estimator (ROSTAT) to derive the cluster redshift and
velocity dispersion (Beers et al., 1990). Even
though limited in size, this spectroscopic galaxy sample is generated by an almost error-free
cluster membership selection process.

\begin{figure}
\includegraphics[width=0.8\linewidth]{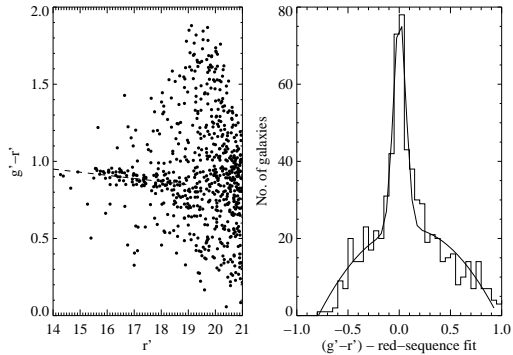}
\caption[cmd3.eps]{\label{fig:cmd}
Sample galaxy cluster color magnitude diagram with fitted red sequence as described in section 2.3}
\end{figure}

In order to increase our sample size we alternatively resort to a less accurate,
photometric, membership selection (sample B). Color-magnitude diagrams (CMDs) for
galaxies in cluster fields allow cluster membership to be assessed
based on galaxy colors. Figure~\ref{fig:cmd} shows the CMD for one
of the clusters analysed in this paper (Abell 1773). The dashed line
indicates the best fit to the red sequence, a prominent feature of the galaxy population of 
massive clusters. The red sequence is a consequence of the $D_{4000}$
break - a characteristic of old stellar populations ubiquitous in elliptical galaxies - and Dressler's morphology-density relation (1980). By choosing filters which straddle
this break at the redshift of the cluster (for this work, g' and r'), elliptical galaxies in the cluster can easily be selected. 

We use an iterative, automatic fitting procedure for delimiting the red-sequence boundaries in order to minimize observer bias.  We collapse all data points along the best linear fit to the red sequence and model the resulting dispersion histogram as a second-order polynomial (field galaxies) with a superimposed Gaussian (red-sequence galaxies), as shown in Figure~\ref{fig:cmd}. We iterate this process until the fit converges, and select cluster members as all galaxies falling within $\pm 2\sigma$ of the red sequence. In order not to miss the cD galaxies, which tend to be bluer than the rest, we also extract any bright objects that are bluer than the red sequence and within 250 kpc of the cluster center. The signal-to-noise ratio for the sample is optimized by taking the extraction radius for each cluster to be the point where the local surface density of red-sequence galaxies in the cluster area equals that in the background region. 

A lower X-ray luminosity cutoff for the host clusters is imposed at
$0.4\times 10^{44}$ ergs s$^{-1}$ ($0.1$ - $2.4$ keV) in order to exclude systems that are too poor to
exhibit a pronounced red sequence in the available data, reducing the final sample to 85 clusters, with redshifts ranging from $z=0.02$ to $z=0.23$.
Finally, a control sample (sample C) is compiled for each cluster from 
all the galaxies within the background region that fall inside the cluster's red-sequence boundaries in color-magnitude space.

\section{Alignment Results}

Following Struble \& Peebles (1985), we describe our results in terms of an alignment parameter $\delta=\sum_i \frac {\phi_i}{N}-45^{\circ}$, where $\phi_i$ are the galaxy orientations with respect to the cluster center ($0^{\circ}$ being the radial direction and  $90^{\circ}$ the tangential one) and $N$ is the number of galaxies in the sample. For an isotropic distribution, $\delta$ should be consistent with zero, whereas radial alignment will lead to a negative value of $\delta$. We determine
$\delta$ for each individual cluster and for samples A, B and C and compute its $1\sigma$ statistical uncertainty (derived for an isotropic distribution), $\sigma_{\delta} = \frac{90}{\sqrt{12 N}}$. For galaxy sample A (2210 spectroscopically selected galaxies in 48 clusters) we find $\delta = -2.21^{\circ} \pm 0.55 ^{\circ}$, i.e. a weak, but statistically significant, preference for radial galaxy
   alignment at the $4\sigma$ confidence level. Galaxy sample B (10,472 galaxies from 85 different clusters) yields only a marginal net alignment of $\delta = -0.48^{\circ} \pm 0.25^{\circ}$. However, when limited to galaxies with $r'<18$ (the approximate limiting magnitude of sample A), this sample shows a much more significant galaxy alignment, characterized by $\delta = -1.06^{\circ} \pm 0.37^{\circ}$ (4,853 galaxies). The control sample C (7,396 galaxies over 85 clusters) yields, as expected, a null result: $\delta = 0.69^{\circ} \pm 0.48^{\circ}$ and $\delta = 0.30^{\circ} \pm 0.30^{\circ}$, with and without the limitation to $r'<18$ galaxies.  

\begin{figure}
\vspace*{3mm}
\hspace*{-2mm}\includegraphics[width=0.8\linewidth]{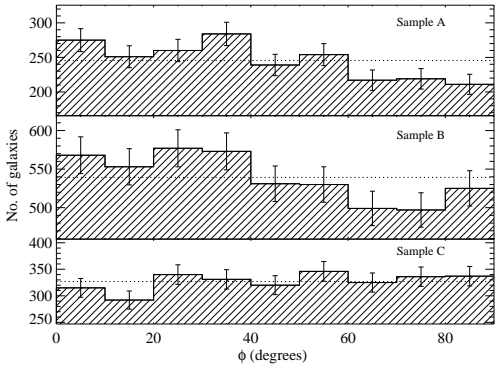}\\*[6mm]
\caption[hist2.eps]{\label{fig:hist}
Galaxy alignment distribution for samples A,B and C. Error bars assume poisson statistics and dotted lines show the expected distribution of a 3-dimensional isotropic sample }
\end{figure}

We test the robustness of these results by 
   comparing the distribution of galaxy orientations in all three samples
   with an isotropic distribution (Fig.~\ref{fig:hist}) and, again, find strong evidence
   for a significant radial alignment of cluster galaxies. A one-sided non-parametric Kolmogorov-Smirnov test yields probabilities of $1.7 \times 10^{-4}$, $3.4 \times 10^{-3}$  and $0.35$, for samples A, B and C, respectively, being drawn from an isotropic distribution.  

In order to verify that our measured radial alignment is not caused exclusively by a preferred galaxy orientation along the cluster axis combined with a flattened galaxy distribution along the same direction,  we measured the global ellipticity, $\epsilon$, of each cluster by computing the
spatial moments of its galaxy distribution. We find no evidence that the alignment signal is dominated by flattened clusters: for round clusters ($\epsilon<0.25$) we find
$\delta = -2.48^\circ\pm 0.77^\circ$, and for flattened clusters ($\epsilon>0.25$) we obtain
$\delta = -1.92^\circ\pm 0.82^\circ$.

\begin{figure}
\includegraphics[width=0.9\linewidth]{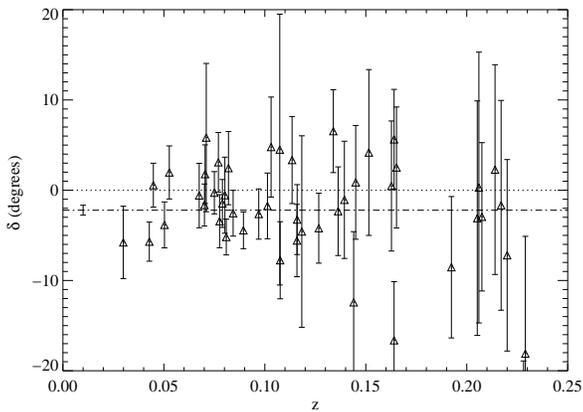}
\caption[z2.eps]{\label{fig:z}
The measured alignment parameters $\delta$ and their errors, $\sigma$
for sample A are plotted versus cluster redshift. The dashed line shows
the sample alignment and its one-sigma error and the dotted line shows
the expected alignment for an isotropic distribution}
\end{figure}

As mentioned in the introduction, any correlations of the strength of this alignment with individual cluster properties will be crucial towards understanding the physical mechanisms behind it. To look for a  correlation with morphology, we used a combination
of X-ray and optical images to separate clusters into 3 morphology
bins, ranging from extremely relaxed systems to bimodal, heavily
substructured clusters. We found no statistically significant
correlation between alignment and morphology or redshift, a result that is perhaps
not unexpected given the statistical errors in our sample. To see this
we plot all of our sample A cluster alignment parameters versus redshift in
Figure~\ref{fig:z}. The dashed line shows the best-estimate
sample alignment parameter and its $1\sigma$ error. We find that the
reduced chi-squared statistic, $\chi^2_r=1.10$, describes a cluster
sample that is characterized by a single value of $\delta$ and which is consistent with our previous error estimates.
We conclude that further investigation into possible correlations of the
alignment effect with individual cluster properties will require
a larger cluster sample and/or deeper optical data. 

\subsection{Systematics}

We are not aware of a systematic effect that can yield a
ficticious tendency for radial alignment in samples A and B, without
also affecting the control sample C.  Nevertheless, we delve briefly here into 
the most obvious sources of error. Instrumentally, these are not pointed observations, but part of a wide-field survey. The position of the cluster centers on the observed fields is therefore random, and we can ignore issues such as PSF anisotropies and array scanning errors.



Errors in the SDSS position angles for faint galaxies determine the limiting 
magnitude we can work with. The SDSS does not provide an error estimate for this quantity, but we find that the variance in the isophotal orientation increases steeply around $r'=19.2$, and adopt this limiting magnitude as an absolute cut-off for the red-sequence selection in samples B and C. As discussed in section 3., including faint galaxies $18<r'<19.2$ in sample B  reduces the significance of our result. We looked for a dependence on galaxy brightness by splitting the galaxies for each sample A cluster evenly into bright and faint subsamples.  We found the alignment tendency to be effectively equal ($\delta_{bright} = -2.10^\circ \pm 0.78^{\circ}$ and $\delta_{faint} = -2.37^\circ \pm 0.78^{\circ}$), and conclude that for $r'\lesssim18$ there is no evidence that the alignment effect is dependent on galaxy brightness. 



\section{Parametric Tidal Effects on Cluster Galaxies}
In this section we will argue for the existence and importance of dynamical tidal effects for cluster galaxies by drawing an analogy between a cluster-galaxy system and the Milky Way (MW)-Dwarf Spheroidal (dSph) systems.  
The dynamical properties of elliptical galaxies in clusters are, in some ways, similar to dSphs in the Local Group.  Fleck and Kuhn (2003-henceforth FK) argued that many of the Local Group
dSph galaxies have been strongly affected by the time-variable MW
tide.  When the internal dSph gravitational timescale is comparable to the dSph-MW
orbit period important resonance effects are likely. The tidal force
parametrically excites the dSph, and the Mathieu equation describing this
interaction exhibits secular growing solutions under a broad range of
orbit and galaxy density conditions. This is a key feature of parametric
oscillations, where resonance depends strongly on the
amplitude of the driving force, and weakly on the commensurability of the
internal and external timescales. FK found that, under
dSph conditions, 
parametric amplification can have a significant influence on the dynamical state of these systems over a Hubble time, provided the MW orbital angular
frequency $\omega = \sqrt{GM(r)\over r^3}$ is within a factor of about
3 of the fundamental gravitational frequency of the stellar system
$\omega_0 = \pi\sqrt{G\rho}$, where $M(r)$ is the enclosed MW mass at
radius $r$.  This condition is satisfied by some of the local dSph, and
FK argued that it generally accounts for their ellipticities, velocity
dispersions, and extended extratidal stellar populations.

Although our SDSS sample spans a wide range of elliptical galaxy densities
and  cluster masses, it is worth noting that a typical elliptical galaxy from this sample and its host cluster tend to satisfy the parametric
resonance conditions. Padmanabhan et al. (2004) used SDSS data to
generate mass models for elliptical galaxies.  Their "typical''
SDSS mass model (their Figure 2.) leads to a pulsational frequency of $\omega_0=2\pi /T$
with $T$=30 Myr. Cluster mass models based on X-ray and galaxy velocity
dispersion constraints (e.g. Waxman and Miralda-Escude 1995) imply
cluster masses of about $10^{14}M_\odot$ at a central distance of
100kpc. This leads to a circular orbit period of 75 Myr which is within
a factor of 2.5 of the galaxy internal timescale.  The
range in masses for elliptical galaxies in these SDSS clusters is typically
two orders of magnitude, and the range in cluster masses is similarly
large. Thus, our sample includes clusters and member galaxies with a
range of pulsational and orbital periods which broadly encompass
parametric resonance conditions.

The importance of the FK mechanism is that it increases the likelihood
of tidal interaction between the cluster gravitational field and its
constituent galaxies. This can have three important effects:  1) it can
create elliptical galaxies within the cluster, 2) it can generate
intracluster light by stripping stars from the galaxies, and 3) it tends
to align the affected galaxies with respect to the cluster
center. Equation (25) in FK describes how galaxies in nearly circular
orbits tend to have stars tidally accelerated away from their
cores along a line that
makes an angle $\theta$ with the direction
to the cluster center. Here $tan\theta \approx
-(1+{\omega\epsilon\over 2(\omega_0-\omega )})$, with $\epsilon$ being a
a dimensionless parameter that expresses
the strength of the cluster gravitational tide. Galaxies affected by
this mechanism will tend to be elliptical and oriented along this direction.
This equation for $\theta$ shows that, as a galaxy orbits in the cluster
potential, it will be elongated along an angle that is greater (less)
than 45$^o$ to the cluster center direction, depending on whether
$\omega_0$ is greater (less) than $\omega$. 
Over time the effect of the parametric resonance is to decrease the galaxy density, so that if $\omega_0$ and $\omega$ were at
any time of comparable magnitude then $\omega_0$ would evolve toward
{\it smaller} values as stars are dynamically extracted from the galaxy
core.  While there will also be a tendency to
circularize the galaxy's orbit as energy goes toward heating
the stellar system, the change in the orbital $\omega$ is smaller
than the density effect on $\omega_0$ and it eventually becomes {\it
less} than $\omega$ even if it was initially larger.  This means that
those galaxies affected by the FK effect in clusters that are at least several crossing times old will tend
to be aligned along the radial direction (having $\theta < 45^o$). FK
also showed numerically that  non-circular orbits exhibit the
same behavior, as they lead to rotating elliptical galaxies with a time
averaged preference for the galaxy to be radially aligned (cf. FK --
Fig. 13). 

It is interesting that a numerical model of cluster galaxies (Muccione and Ciotti 2004) has found that under some conditions 10\% or more of the total galaxy mass can be extracted into the intracluster environment over a Hubble time through what they call ``collisionless stellar evaporation''. Because they were simulating galaxies with stellar crossing times comparable to cluster orbital periods, it is likely that they are also seeing this aspect of the parametric tidal mechanism.

We lack sufficiently detailed cluster galaxy orbit and central density
information to predict how any given elliptical galaxy should be
aligned.  On the other hand we do expect a statistical tendency toward radial alignment of cluster galaxies. This is reminiscent of the
dissipative dynamic tidal locking mechanisms that cause planetary spin-orbit coupling in solar system objects.


\section{Discussion}

Our analysis suggests that galaxies show a statistically robust tendency toward radial
alignment in low-redshift clusters ($z \le 0.23$).  While there may be variations in the strength of the
effect between different clusters, to our measurement sensitivity, the observed radial alignment appears to be a uniform and robust feature of the 85 X-ray selected
clusters we studied using the SDSS dataset. The detection of such a net alignment of galaxies in a low-redshift cluster sample seems to point to the existence of a mechanism, such as the dynamical tidal interaction described above, that can increase the net alignment over time so that it remains observable in the nearby universe. However, it is also possible that what we are seeing are remnants of a primordial "imprint" that has not yet been washed away by the tumultuous events in a cluster's history. It is vital to expand this analysis to cover clusters at higher redshifts and systems with
   different formation histories so that the effect's evolution with time can be studied and the two scenarios can be distinguished. With a more sensitive study of the correlations of galaxy
alignment with projected orbit, galaxy type, and cluster
morphology, we look forward to using alignment as a tool to probe, and
perhaps distinguish between, possible mechanisms behind elliptical galaxy and
cluster formation and evolution.

\acknowledgements

We thank Harald Ebeling for helpful advice on the eBCS, X-ray data analysis, and morphology classifications.


\end{document}